\documentclass[aps,prl,reprint,amsmath,amssymb]{revtex4-2}

\usepackage{hyperref}
\usepackage{graphicx}
\usepackage{xcolor}

\renewcommand{\vec}[1]{\boldsymbol{#1}}
\newcommand{\p}[0]{\partial}

\newcommand{\bmth}[1]{\mbox{\boldmath $#1$}}
\newcommand{\grad}{\bmth{\nabla}}
\newcommand{\Ca}{\mathrm{Ca}}
\newcommand{\C}{\mathcal{C}}
\newcommand{\res}{\mathrm{res}}
\newcommand{\atm}{\mathrm{atm}}

\begin{document}
	\title{Gas compression systematically delays the onset of viscous fingering}
	\author{Liam C. Morrow}\thanks{These authors contributed equally to this work.}
	\author{Callum Cuttle}\thanks{These authors contributed equally to this work.}
	\author{Christopher W. MacMinn}
	\email{christopher.macminn@eng.ox.ac.uk}
	\affiliation{Department of Engineering Science,\\
		University of Oxford, Oxford, OX1 3PJ, UK \\
	}
	
\begin{abstract}
	Using gas to drive liquid from a Hele-Shaw cell leads to classical viscous fingering. Strategies for suppressing fingering have received substantial attention. For steady injection of an incompressible gas, the intensity of fingering is controlled by the capillary number $\Ca$. Here, we show that gas compression leads to an unsteady injection rate controlled primarily by a dimensionless compressibility number $\C$. Increasing $\C$ systematically delays the onset of fingering at high $\Ca$, highlighting compressibility as an overlooked but fundamental aspect of gas-driven fingering.
\end{abstract}

\maketitle

When a viscous liquid is displaced from a Hele-Shaw cell or porous medium by the injection of a gas, the gas-liquid interface is hydrodynamically unstable and tends to deform into branched, finger-like structures~\citep{Hill1952,Saffman1958,Chuoke1959,Homsy1987}. This viscous-fingering instability has been extensively studied as an archetype of interfacial pattern formation~\citep{Paterson1981,McCloud1995,Couder2000,Casademunt2004} and for its relevance to a variety of practical applications, including the operation of fuel cells~\citep{Lee2019,Mortazavi2020}, the remediation of groundwater contamination~\citep{Clayton1998,Hu2010}, and the subsurface sequestration of CO$_2$~\citep{Wang2013,Chen2017} or storage of hydrogen~\citep{Paterson1983}. 
A concern in all applications is understanding the key mechanisms that promote or suppress the development of this instability, which is driven by viscous forces in the defending phase and opposed by capillary forces at the interface. The capillary number $\Ca$ compares the magnitudes of these two forces and is thus the key parameter in determining the intensity of fingering. Recent decades have seen an increased focus on exploring the impact of perturbations to the flow or to the flow cell at a given $\Ca$, including active modulation of the injection rate~\citep{Dias2012, Morrow2019} and use of a geometry that varies in space~\cite{zhao-pra-1992, dias-pre-2010b, AlHousseiny2012,dias-pre-2013, rabbani-pnas-2018, Morrow2019} and/or in time \citep{PihlerPuzovic2012, al-housseiny-prl-2013, Zheng2015, Morrow2019, anjos-pre-2021}.

One fundamental aspect of this problem that has thus far been ignored is the volumetric compression of the injected gas, which is practically unavoidable under the typical pressures associated with displacing a viscous liquid from a confined geometry. Previous theoretical studies of fingering have taken both fluids to be incompressible and many previous experimental studies have implicitly made the same assumption. Gas compression can lead to unsteady flow or even stick-slip motion, thereby exerting a fundamental influence on displacement processes and pattern formation~\citep{Sandnes2011,Sandnes2012, Lai2018, Cuttle2023a}, but the implications of compression on viscous fingering have been overlooked. Here, we use mathematical modelling, simulations, and experiments to study viscous fingering driven by the steady compression of a gas reservoir. We find that steady compression leads to an unsteady injection rate that is dictated by a dimensionless compressibility number $\C$, which can be interpreted as the rate of viscous depressurisation relative to the rate of compressive pressurisation~\citep{Cuttle2023a}. We find that increasing $\C$ delays the onset of the instability to a degree that is comparable to that of decreasing $\Ca$ by a similar magnitude. Our results suggest that compression can play a fundamental role in viscous fingering, with implications for many previous studies of interfacial instabilities during gas-driven displacement.

\begin{figure*}
\includegraphics[width=1.0\linewidth]{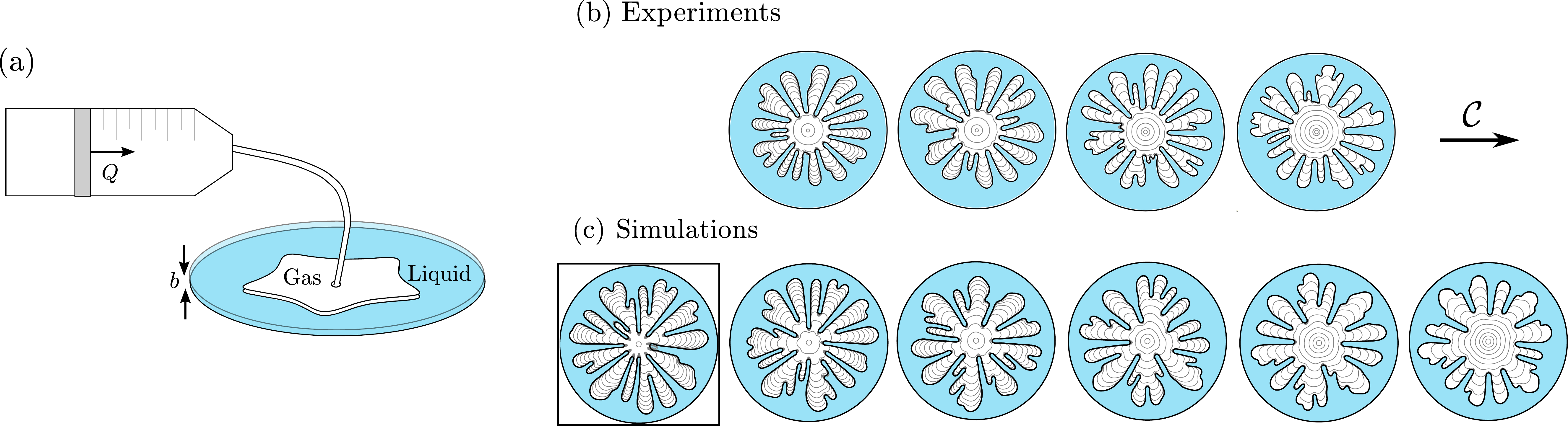}
\caption{(a) A body of gas with initial volume $V_g(0)$ is compressed at a steady volumetric rate $Q$ in order to displace liquid from the gap $b$ between two plates, leading to viscous fingering. (b) Experiments performed with $Q = 2.5$~mL/min for (left to right) $V_g(0) = 25$, $50$, $100$, and 200~mL. Other parameters are $b = 420$ $\mu$m, $R_c$ = 10.5 cm, $\mu = 0.97$ Pa $\cdot$ s, and $\gamma = 2.1 \times 10^{-2}$ N/m, such that $\Ca = 5.21\times10^3$ and $\alpha = 250$ in all cases.  (c) Corresponding fingering simulations [\textit{i.e.}, numerical solutions to Eqs.~\eqref{eq:Model1}-\eqref{eq:Model7}]. Column one shows the incompressible case (black box); columns two to six show the compressible case with $p_\atm = 1.01 \times 10^{5}$ Pa and $V_g(0) = 25$, $50$, $100$, $200$, and $400$ mL. For both experiments and simulations, each snapshot shows interface contours at equal time intervals of $\hat{t}_{0.9}/10$. The arrow indicates increasing compressibility number $\mathcal{C}$, which is proportional to $V_g(0)$ and thus increases from left to right.}
\label{fig:figure1}
\end{figure*}

We consider a radial Hele-Shaw cell comprising a narrow gap of thickness $b$ between two rigid, circular plates of radius $R_c$ [Fig.~\ref{fig:figure1}(a)]. The gap is initially filled with an incompressible liquid of viscosity $\mu$ and surface tension $\gamma$. This liquid is displaced by injecting gas through a hole in the centre of one plate. Gas injection is driven by the steady compression of a connected gas reservoir at volumetric rate $Q$, such that the volume of the reservoir is $V_{\res}(t) = V_{\res}(0) - Q t$. Figure~\ref{fig:figure1}(b) shows the time-evolution of the gas-liquid interface in experiments, where silicone oil (1,000~cSt, Sigma) was displaced by air compressed with a syringe pump. These images suggest that increasing the initial reservoir volume $V_{\res}(0)$ (left to right) while holding all other parameters constant systematically delays the onset of fingering, in the sense that the interface remains near-circular up to larger radii. We investigate these observations quantitatively below using full numerical simulations of viscous fingering, complemented by experiments and by a simplified axisymmetric model. We observe strong qualitative and often quantitative agreement between experiments and simulations, as illustrated in Figure~\ref{fig:figure1}(c). We compare our experiments and simulations in detail alongside further analysis of the axisymmetric model in a companion study~\citep{Cuttle2023b}. Here, we focus on the impact of compressibility on the fingering pattern.

To simulate viscous fingering, we model flow of the liquid as Stokes flow averaged over the gap $b$. We make the problem dimensionless via
\begin{equation}
\begin{split}
	\hat{\vec{x}} = \frac{\vec{x}}{R_c}, \quad & \hat{t} = \frac{Q}{\pi R_c^2 b}t, \quad \hat{p} = \frac{\pi b^3}{6 \mu Q} p, \\ &\hat{\vec{v}} = \frac{2\pi R_c b}{Q} \vec{v}, \quad \hat{V} = \frac{V}{\pi R_c^2 b},
\end{split}
\end{equation}
leading to the following dimensionless parameters:
\begin{equation}
\begin{split}
	\Ca = \frac{12 \mu Q R_c}{2\pi b^3 \gamma}, \quad \C = \frac{12 \mu Q V_g(0)}{\pi^2 R_c^2 b^4 p_{\atm}},  \quad \mathcal{V} = \frac{V_g(0)}{\pi R_c^2 b},  \\ \alpha = \frac{R_c}{b}, \quad \mathcal{R} =\frac{r_0}{R_c},
\end{split}
\end{equation}
where the total volume of gas $V_g = V_{\res} + V_{b}$ combines the volume $V_{\res}(t)$ of gas in the reservoir and the volume $V_{b}(t)$ of gas in the cell. The capillary number $\Ca$, commonly referred to as the ``modified'' capillary number, incorporates the role of the aspect ratio $\alpha$. The new compressibility number $\C$ compares viscous and compressive (atmospheric) pressure scales or, equivalently, the rates at which these pressures vary. We fix the geometry of the flow cell and the radius $r_0$ of the initially circular bubble, such that $\alpha = 250$ and $\mathcal{R}=0.025$. The nondimensional flow equations are then
\begin{alignat}{3}
	\hat{\grad}^2 \hat{p} &= 0 \quad &\textrm{in } &\hat{\vec{x}} \in \mathbb{R}^2 \backslash \Omega, \label{eq:Model1} \\
	\hat{v}_n &= - \frac{1}{1 - f_1}  (\hat{\grad} \hat{p}) \cdot \vec{n} \quad &\textrm{on } &\p \Omega, \label{eq:Model2}\\
	\Delta\hat{p} &= \Delta\hat{p}_g(\hat{t}) - \left( \frac{\pi \hat{\kappa}}{4} + 2\alpha f_2 \right) \quad &\textrm{on } &\p \Omega, \label{eq:Model3} \\
	\Delta\hat{p} &=0\quad &\textrm{on } &|\hat{\vec{x}}| \in 1, \label{eq:Model4}
\end{alignat}
where $\p{\Omega}$ denotes the gas-liquid interface, $\hat{v}_n$ is the magnitude of the normal velocity of the interface, $\vec{n}$ is the unit normal to the interface, and $\hat{\kappa}$ is the signed in-plane curvature of the interface. Gauge pressures of the liquid and gas, $\Delta p$ and $\Delta p_g$, respectively, are measured relative to $p_\atm$. To capture the effects of thin residual films of the wetting liquid in the gas region \citep{Park1984}, the functions $f_1$ and $f_2$ modify the kinematic and dynamic boundary conditions, respectively, to account for volume conservation, and enhanced curvature and viscous stresses at the interface. Following \citet{Peng2015}, we adopt the empirical expressions
\begin{align}
f_1 &= \frac{\left| \mu v_n/\gamma \right|^{2/3}}{0.76 + 2.16\left| \mu v_n/\gamma \right|^{2/3}}, \label{eq:Model5} \\
f_2 &= 1 + 1.59 \left| \mu v_n/\gamma  \right|  + \frac{\left| \mu v_n/\gamma  \right|^{2/3}}{0.26 + 1.48\left| \mu v_n/\gamma  \right|^{2/3}}, \label{eq:Model6}
\end{align}
which were derived by fitting to simulations of viscous fingering in a rigid Hele-Shaw channel~\cite{Reinelt1985}.

We ignore pressure gradients in the gas, taking $p_g(t)$ to be spatially uniform. We also assume isothermal compression, such that $p_g(t) = p_g(0) V_{g}(0) / V_{g}(t)$, which is justified by our experimental observations~(see Appendix~A of Ref.~\citep{Cuttle2023b}). The dimensionless gauge gas pressure may then be written
	\begin{equation}
		\Delta\hat{p}_g = \frac{ 2\C^{-1}\left(\hat{t}+\mathcal{R}^2-\hat{V}_b\right) + \Ca^{-1} \left(\frac{\pi}{4\mathcal{R}}+2\alpha\right) }{1 - \mathcal{V}^{-1} \left(\hat{t} + \mathcal{R}^2 - \hat{V}_{b} \right)}.   \label{eq:Model7}
	\end{equation}

In addition to the above compressible model, we also consider an analogous incompressible model, in which Eqs.~\eqref{eq:Model1}-\eqref{eq:Model6} are unchanged but $\Delta\hat{p}_g(\hat{t})$ is no longer governed by Eq.~\eqref{eq:Model7}, instead evolving in time such that $\hat{Q}_b \equiv \mathrm{d}{\hat{V}_b}/\mathrm{d}\hat{t} = 1$ (\textit{i.e.}, $Q_b\equiv\mathrm{d}{V}/\mathrm{d}{t}=Q$).

We solve both compressible and incompressible models numerically using a scheme proposed by \citet{Morrow2021}, which utilises the level-set method \citep{Osher1988}. Our simulations use the perturbed initial condition 
\begin{align}
\hat{R}(\theta, 0) = \mathcal{R} \left\{ 1 + \varepsilon \sum_{n=2}^{12} \cos \left[ n (\theta - 2 \pi r_n) \right] \right\}, \label{eq:InitialCondition}
\end{align}
where $\hat{R}(\theta,\hat{t})$ is the radial position of the interface at azimuthal coordinate $\theta$, and $\varepsilon \ll 1$ and $0 \le r_n \le 1$ are random numbers sampled from uniform distributions. Simulations are performed on the domain $0 \le \hat{r} \le 1$ and $0 \le \theta < 2 \pi$ using 1,000 $\times$ 1,000 equally-spaced nodes, and are concluded when $\hat{R}\ge 0.99$ for any value of $\theta$. 

To better understand the coupling between gas compression and fluid displacement, we also consider an axisymmetric model in which the interface is taken to be a circle of radius $\hat{R}_0\left(\hat{t}\right)$. By further taking the limits that the air reservoir is much larger than the volume of the Hele-Shaw cell ($\mathcal{V} \gg 1$) and that the capillary pressure is negligible relative to atmospheric pressure ($2\C^{-1} \gg \Ca^{-1} [\pi/(4\mathcal{R})+2\alpha]$), and ignoring thin-film effects ($f_1 =0$, $f_2 = 1$), Eqs.~\eqref{eq:Model1}-\eqref{eq:Model7} reduce to
\begin{align}
\frac{\textrm{d}\hat{R}_0}{\textrm{d}\hat{t}} = \frac{\Delta \hat{p}_g}{2\hat{R}_0 \ln \left(1/\hat{R}_0\right)}, \quad \Delta \hat{p}_g = \frac{2(\hat{t} + \mathcal{R}^2 - \hat{R}_0^2 )}{\mathcal{C}}. \label{eq:RadSym}
\end{align}
The dynamics of this axisymmetric model [Eq.~\eqref{eq:RadSym}] are then determined solely by the compressibility number. For the incompressible case ($\C = 0$), the solution to the axisymmetric model is $\hat{R}_0 = \sqrt{\mathcal{R}^2 + \hat{t}}$ and $\Delta\hat{p}_g=\ln(1/\hat{R}_0)$.

\begin{figure}
\centering
\includegraphics[width=1\linewidth]{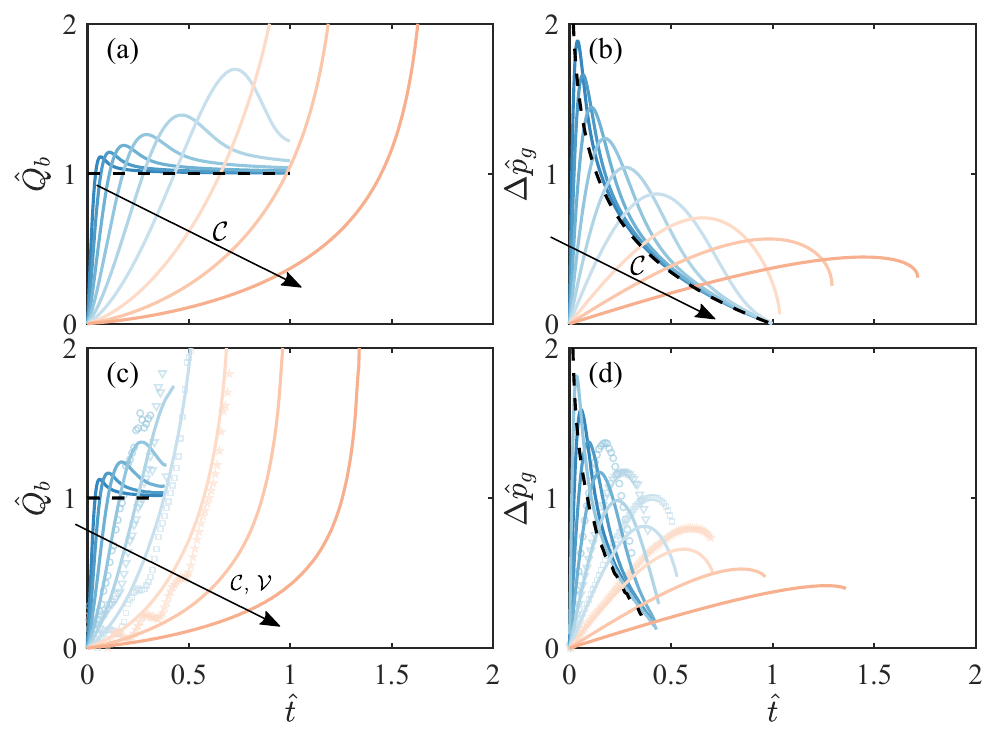} 
\caption{Nondimensional (left) injection rate $\hat{Q}_b$ and (right) gauge gas pressure $\Delta \hat{p}_g$ for $\mathcal{C} = 0.018$, $0.036$, $0.071$, $0.14$, $0.28$, $0.57$, $1.13$, $2.27$, and $4.54$ (dark blue to dark red; arrows). Blue and red curves are for $\mathcal{C}<1$ and $\mathcal{C}>1$, respectively. The incompressible case is also shown (dashed black). Panels (a) and (b) are from the axisymmetric model [Eq.~\eqref{eq:RadSym}]. Panels (c) and (d) are from fingering simulations (solid curves) and experiments (symbols), both for $\Ca = 2.08 \times 10^{4}$ with $\mathcal{V}=0.215, 0.430, 0.859, \textbf{1.72, 3.44, 6.87, 13.7}, 27.5,$ and 55 (experimental values in bold).}
\label{fig:VolumePressure}
\end{figure}

We solve Eq.~\eqref{eq:RadSym} numerically to plot the actual, time-varying injection rate $\hat{Q}_b\left(\hat{t}\right)=\textrm{d}\hat{V}_b/\textrm{d}\hat{t} = 2 \hat{R}_0 (\textrm{d}\hat{R}_0/\textrm{d}\hat{t})$ and the gauge gas pressure in Figs.~\ref{fig:VolumePressure}(a,b), respectively, for a range of $\mathcal{C}$. Both quantities are initially zero for all $\mathcal{C} > 0$. Both then increase toward and then beyond the incompressible solution, corresponding to an initial period of increasing injection rate due to gas pressurisation. The pressure evolves non-monotonically in time for all $\C$, increasing at early times when the compression rate exceeds the injection rate ($\hat{Q}_b<1$) and then decreasing at late times, once the injection rate exceeds the compression rate ($\hat{Q}_b>1$). The injection rate $\hat{Q}_b$ is determined by a balance between the gauge gas pressure and the viscous resistance in the draining liquid. This observation can be used to demonstrate the existence of two dynamical regimes in the axisymmetric model~\citep[see][]{Cuttle2023b}. For $\C \lessapprox 1$ (blue curves), the breakout time $\hat{t}_f\equiv\hat{t}(\hat{R}_0=1)$ is unity and the breakout pressure $\Delta\hat{p}_g(\hat{R}_0=1)$ is zero, corresponding to a nonmonotonic evolution of $\hat{Q}_b$, which tends to a finite value at breakout. For $\C \gtrapprox 1$ (red curves), breakout is delayed ($\hat{t}_f>1$) and the breakout pressure is significantly greater than zero. This overpressure leads to a monotonically increasing and ultimately divergent injection rate as the viscous resistance vanishes at breakout. These regimes arise in the axisymmetric model due to the coupling of time-evolving compressive and viscous forces, independent of viscous fingering. The same features were recently identified and analysed in detail in the context of a capillary tube~\citep{Cuttle2023a}.

Figures~\ref{fig:VolumePressure}{(c, d)} show $\hat{Q}_b$ and $\Delta \hat{p}_g$, respectively, from fingering simulations at the same values of $\mathcal{C}$ as in Figs.~\ref{fig:VolumePressure}(a, b) and for $\Ca = 2.08 \times 10^4$, demonstrating that the qualitative dynamics of the axisymmetric model and the fingering simulations are strikingly similar. The presence of viscous fingering, and to a lesser extent residual films, leads to a few key differences. Notably, the fingering simulations achieve breakout much earlier (typically around $\hat{t}_f\approx0.5$ for low $\mathcal{C} \lessapprox 1$) because the irregular, branched interface bypasses a substantial fraction of the defending liquid. In addition, the transition to monotonically increasing $\hat{Q}_b$ occurs significantly below the threshold $\mathcal{C}\approx 1$ from the axisymmetric model. (Note that the threshold $\mathcal{C}$ is approximate in the Hele-Shaw geometry because of a secondary dependence on $\mathcal{R}$ \citep{Cuttle2023b}.) Unlike in the axisymmetric model, the values of $\hat{Q}_b$ and $\Delta \hat{p}_g$ from the fingering simulations do also depend on the value of $\Ca$, but we find that this dependence is comparatively weak, as discussed below. Nonetheless, this comparison suggests that the basic physics of compression-driven displacement underpin the dynamics of this system, even in the presence of viscous fingering. For comparison, we also plot experimental measurements at four values of $\C$ in Fig.~\ref{fig:VolumePressure}{(c, d)}, which show excellent qualitative agreement with the fingering simulations (see Ref.~\citep{Cuttle2023b} for a detailed comparison).

\begin{figure}
\centering
\includegraphics[width=0.99\linewidth]{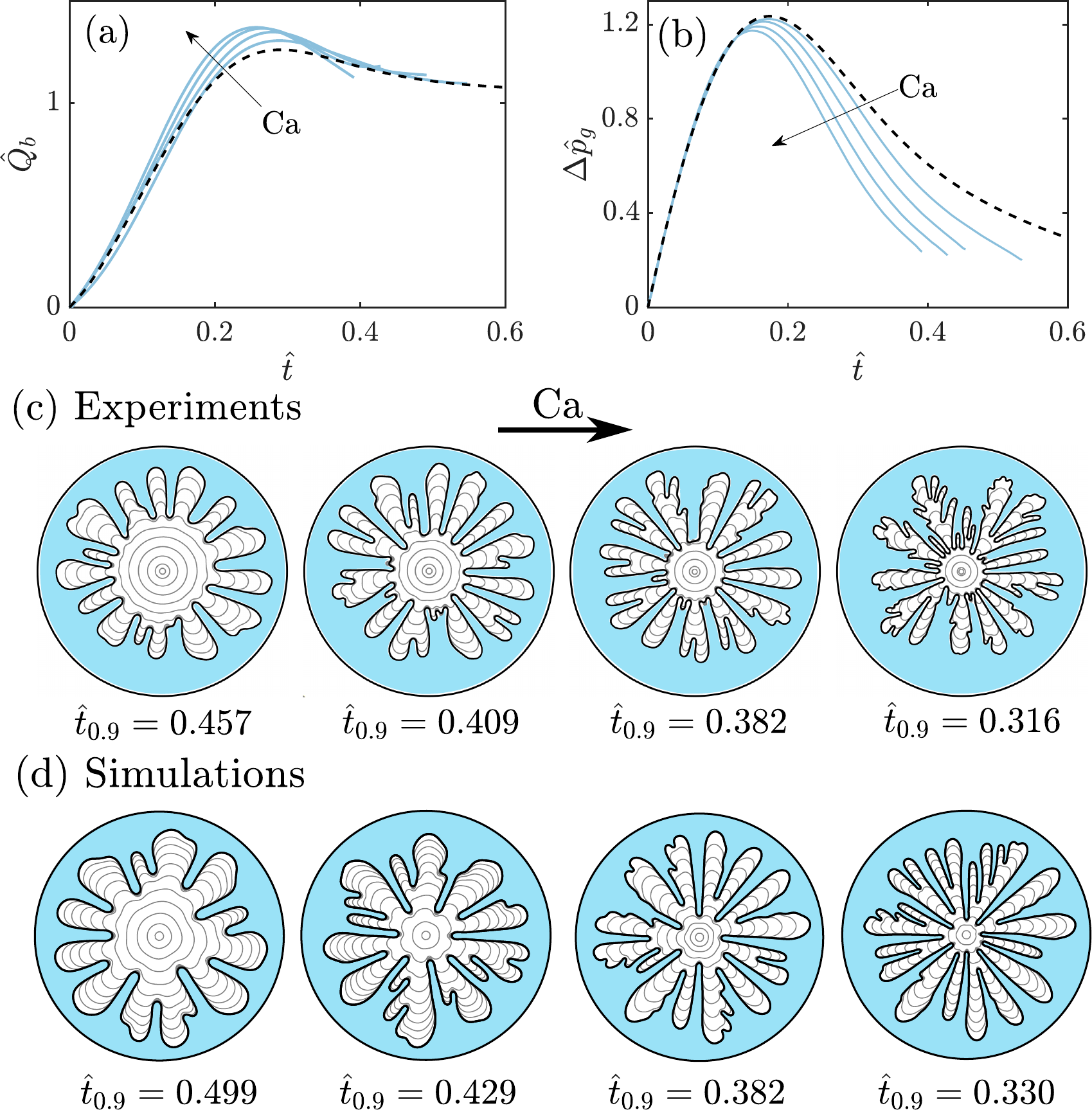}
\caption{Nondimensional (a) injection rate $\hat{Q}_b$ and (b) gauge gas pressure $\Delta \hat{p}_g$ from fingering simulations for $\mathcal{C} = 0.14$ and $\Ca=2.61\times10^3$, $5.21\times10^3$, $1.04\times10^4$, and $2.08\times10^4$ (solid blue curves; arrows show increasing $\Ca$). The solution to the axisymmetric model [Eq.~\eqref{eq:RadSym}] is shown for comparison (dashed black). Panels (c) and (d) show the evolution of the fingering pattern from corresponding experiments and simulations, respectively, plotted at equal time intervals of $\hat{t}_{0.9}/10$.}
\label{fig:VaryCa}
\end{figure}

The importance of $\C$ is further illustrated by considering fingering simulations at fixed $\mathcal{C}$ and varying $\Ca$ as plotted in Figs.~\ref{fig:VaryCa}(a, b). In practice, this can be achieved by varying $Q$ while holding $QV_g(0)$ and all other parameters constant. The results of the axisymmetric model are included for reference (dashed black). The injection rate $\hat{Q}_b$ [Fig.~\ref{fig:VaryCa}(a)] is particularly insensitive to variations in $\Ca$ and, hence, to the evolution of the fingering pattern. The gauge gas pressure $\Delta \hat{p}_g$ [Fig.~\ref{fig:VaryCa}(b)] appears to be somewhat more sensitive to the pattern, featuring a more pronounced $\Ca$-dependent deviation from the axisymmetric solution after the peak pressure is reached. However, variations in $\Ca$ clearly have much less of an impact on $\hat{Q}_b$ and $\Delta \hat{p}_g$ than variations in $\mathcal{C}$ of a similar magnitude [cf. Figs.~\ref{fig:VolumePressure}(c, d)]. This insensitivity to $\Ca$ is surprising, given the strong impact of $\Ca$ on the fingering pattern, as illustrated in Fig.~\ref{fig:VaryCa}(c, d) for the corresponding experiments and simulations, respectively. The severity of the fingering pattern increases with $\Ca$, with the instability beginning at smaller radii, generating narrower and more-branched fingers, and leading to earlier breakout---see the values of the near-breakout time $\hat{t}_{0.9}$ at which the interface first crosses $\hat{R}=0.9$, in Figs.~\ref{fig:VaryCa}(c, d). However, the evolving balance between compressive and viscous forces is a bulk feature that is relatively insensitive to the shape of the interface.

\begin{figure}
\centering
\includegraphics[width=1\linewidth]{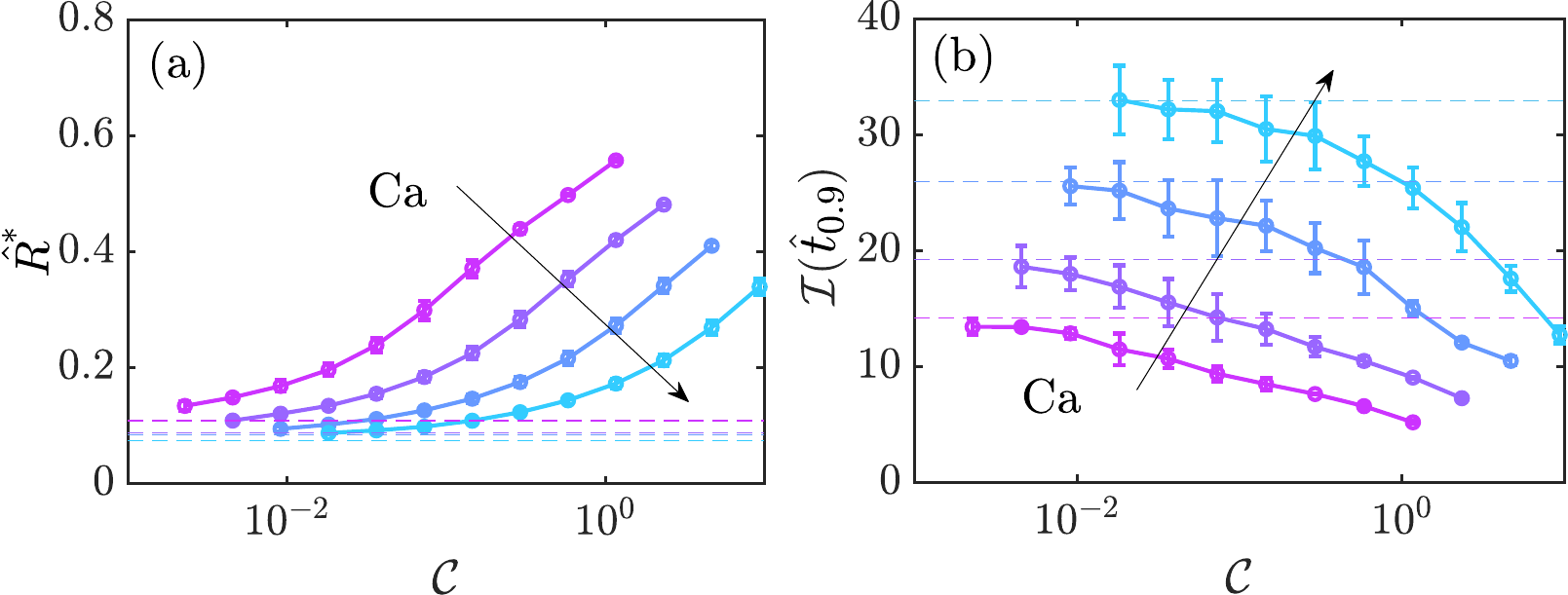} 
\caption{(a) The onset radius $\hat{R}^*$ at which fingers begin to develop, defined as the maximum radius of the interface when $\mathcal{I} = 1.1$, from fingering simuations at varying $\C$ with $\mathcal{V}=$ 0.215, 0.430, 0.859, 1.72, 3.44, 6.87, 13.7, 27.5, 55, and 110. (b)~The subsequent values of $\mathcal{I}$ at the near-breakout time $\hat{t}_{0.9}$. Dashed lines denote the corresponding incompressible cases. Arrows indicate the direction of increasing $\Ca=2.61\times10^3, 5.21\times10^3, 1.04\times10^4$, and $2.08\times10^4$. \label{fig:Isoperimetric} }
\end{figure}

To further examine the influence of $\mathcal{C}$ and $\Ca$ on viscous fingering, we perform a linear stability analysis of the full model [Eqs.~\eqref{eq:Model1}-\eqref{eq:Model7}] by considering the growth of perturbations to a near-circular interface $\hat{R}(\theta, \hat{t}) = \hat{R}_0(\hat{t}) + \varepsilon \gamma_n(\hat{t}) \cos (n \theta) + \mathcal{O}(\varepsilon^2)$, where $\varepsilon \ll 1$, and $\gamma_n$ is the amplitude of mode $n$. This analysis adopts the same assumptions as the axisymmetric model: $\mathcal{V}\gg1$, capillary pressure is negligible compared with atmospheric pressure (but not negligible overall), and we ignore thin films. As such, the base state $\hat{R}_0(\hat{t})$ is determined by Eq.~\eqref{eq:RadSym}, but we retain terms with $\Ca$ in the stability analysis. Following the methodology of \citet{Paterson1981}, it can be shown that the growth rate of each mode is given by
\begin{align}
\lambda_n = \frac{\dot{\gamma}_n}{\gamma_n} = \frac{n-1}{2\hat{R}_0^2} \left[ \hat{Q}_b - \frac{n(n+1)}{\Ca \hat{R}_0} \right], \label{eq:UnstablePerturbation}
\end{align}
where $\hat{Q}_b=2 \hat{R}_0(\mathrm{d}\hat{R}_0/\mathrm{d}\hat{t})$ for a circular interface and $\mathrm{d}\hat{R}_0/\mathrm{d}\hat{t}$ is determined by Eq.~\eqref{eq:RadSym}. The most unstable mode is then
\begin{align}
n_{\max} = \sqrt{\frac{1 + \Ca \hat{R}_0 \hat{Q}_b}{3}}, \label{eq:UnstableMode}
\end{align}
which comes about by solving $\p \lambda_n / \p n = 0$ for $n$. Increasing $\Ca$, $\hat{R}_0$, and $\hat{Q}_b$ all increase $\lambda_n$ and $n_{\max}$, thus promoting the instability (i.e., perturbations grow faster and generate more fingers). In the incompressible system, for which Eqs.~\eqref{eq:UnstablePerturbation} and \eqref{eq:UnstableMode} were originally derived and where $\hat{Q}_b = 1$ and $\hat{R}_0 = (\mathcal{R}^2 + \hat{t})^{1/2}$, $\Ca$ is the key parameter in determining the onset of viscous fingering. For the compressible case, however, both $\hat{Q}_b(\hat{t})$ and $\hat{R}_0(\hat{t})$ depend primarily on $\mathcal{C}$. The very low initial injection rates $\hat{Q}_b\ll1$ observed the axisymmetric model, the fingering simulations, and the experiments [Fig.~\ref{fig:VolumePressure}(a, c)] tend to reduce $\lambda_n$ and $n_{\max}$ considerably at early times relative to an incompressible flow, suggesting that fingering is delayed by compressibility~\citep[see][]{Cuttle2023b}.

To compare the predictions of this linear stability analysis with the nonlinear evolution of the instability in the fingering simulations, we quantify the severity of the fingering pattern using the isoperimetric ratio
\begin{align}
\mathcal{I} \left( \hat{t} \right)  = \frac{\hat{L}^2\left( \hat{t} \right)}{4 \pi \hat{A} \left( \hat{t} \right)}, \label{eq:Isoperimetric}
\end{align}
where $\hat{L}$ and $\hat{A}$ denote the perimeter and area of the gas region, respectively. For a circular interface, $\mathcal{I} = 1$; any deformation away from a circle results in $\mathcal{I}>1$. Hence, the initial deviation of $\mathcal{I}$ away from 1 is an indicator of the onset of instability. Figure~\ref{fig:Isoperimetric}(a) shows the onset radius $\hat{R}^*$ at which fingers begin to develop (defined here as $\max(\hat{R})$ when $\mathcal{I} = 1.1$) as a function of $\mathcal{C}$ for different $\Ca$. As $\mathcal{C} \to 0$, $\hat{R}^*$ asymptotes to the incompressible case. Increasing $\C$ or decreasing $\Ca$ systematically delays the onset of fingering relative to the incompressible case. Varying either $\Ca$ or $\C$ by a similar magnitude has a comparable effect, highlighting the power of $\C$ as a control parameter. While compressibility delays onset primarily by reducing the initial injection rates at higher $\C$, the overshoot in $\hat{Q}_b$ at later times could drive anomalously fast growth and exacerbate the final severity of fingering. As shown in Fig.~\ref{fig:Isoperimetric}(b), however, the delayed onset with increasing $\C$ also correlates with lower $\mathcal{I}(\hat{t}_{0.9})$, meaning a less severe fingering pattern near breakout. Hence, $\C$ not only delays the onset of the instability, but also hinders the nonlinear growth of the fingering pattern. We observe the same qualitative behaviour in experiments over a comparable range of $\C\sim10^{-2}$ -- $10^0$~\citep[see][for a discussion of the distinct patterns formed in experiments and simulations]{Cuttle2023b}.

Thus, compressibility naturally and passively leads to an unsteady injection rate that can significantly delay the onset of fingering and reduce the severity of the fingering pattern, with the compressibility number $\C$ having a comparable effect to that of $\Ca$, the traditional control parameter. Even for relatively small $\C$, which would be unavoidable even in systems that are nominally incompressible, the onset radius is consistently larger and the final isoperimetric ratio is consistently smaller than for the incompressible case [Fig.~\ref{fig:Isoperimetric}], suggesting that compressibility is rarely truly negligible. Our study is of particular relevance to flows in porous media where the presence of compressibility is already widely appreciated in sub-surface fluid injection~\citep{Wang2000}, although its full implications for fluid displacement and fingering were previously not. There is broad scope for taking advantage of compressibility in gas-driven flows in porous media and our study provides a strong foundation for doing so. Further, our results imply that features in other physical systems that play a role analogous to compressibility, such as elasticity, heat capacity, or electrical capacitance, can play an important and even controlling role in the evolution of those systems.

\textit{Data availability} --- The supporting data for this study are openly available on Zenodo \citep{cuttle-zenodo-2023}.

\textit{Published version} --- This article is published as L. C. Morrow, C. Cuttle, and C. W. MacMinn. Gas compression systematically delays the onset of viscous fingering. \emph{Physical Review Letters}, 131:224002, 2023.

\begin{acknowledgments}
	We are grateful to Mr.~Clive Baker for technical support. This work was supported by the European Research Council (ERC) under the European Union’s Horizon 2020 Programme [Grant No. 805469], by the UK Engineering and Physical Sciences Research Council (EPSRC) [Grant No. EP/S034587/1], and by the John Fell Oxford University Press Research Fund [Grant No. 132/012].
\end{acknowledgments}


%

\end{document}